\documentclass[aps,prl,preprint,groupedaddress,nofootinbib]{revtex4-1}
\usepackage{graphicx}
\begin{document}


\title{\Large  Affleck-Dine Baryogenesis with Observable Neutron-Anti-Neutron Oscillation }


\author{\bf Rabindra N. Mohapatra$^a$}
\author{Nobuchika Okada$^b$}
\affiliation{}
\affiliation{$^a$ Maryland Center for Fundamental Physics and Department of Physics, University of Maryland, College Park, Maryland 20742, USA}
\affiliation{$^b$ Department of Physics, University of Alabama, Tuscaloosa, Alabama 35487, USA}


\date{\today}

\begin{abstract}  
We discuss the implications of  Affleck-Dine (AD) baryogenesis  for different classes of baryon and lepton number violating processes: 
specially focussing on implications for neutron-anti-neutron ($n-\bar{n}$) oscillation. 
The class of AD baryogenesis scenarios we work with uses the AD field also as the inflaton which is nonminimally coupled to gravity. 
We find that adequate baryogenesis and no washout by the baryon number ($B$) or the lepton number ($L$) violating operators implies constraints on the observability of the process or in the case of neutrino mass with compatibility with neutrino oscillation observations. In  particular for $n-\bar{n}$ oscillation, we study some of the familiar  operators that connect the AD field to $n-\bar{n}$ oscillation
and find that a split scalar spectrum model turns out to be most advantageous for obtaining an observable $n-\bar{n}$
while remaining consistent with AD baryogenesis. 
It is interesting that this spectrum is similar to a non-supersymmetyric SO(10) model for observable $n-\bar{n}$ oscillation discussed before, suggesting that this AD scenario can be embedded into a grand unified SO(10) model. We also find that for a low scale (all scales in the 100 TeV range), there is a narrow range of parameters where the observable $n-\bar{n}$ oscillation is compatible with viable AD baryogenesis.
A feature of this baryogenesis scenario for $n-\bar{n}$ oscillation is that it necessarily predicts processes with $\Delta B=4$ or higher, all be it with highly suppressed amplitudes.

\end{abstract}

\maketitle

\section{1. Introduction}  
Either baryon number ($B$) or lepton number ($L$) violation is known to be one of the key ingredients in resolving a fundamental puzzle of cosmology, the origin of matter-anti-matter asymmetry. 
Even though the standard model (SM) at the non-perturbative level does have baryon and lepton number violation,
it is too weak to explain the origin of matter by itself and new physics beyond the standard model (BSM) is called for. 
This has not only inspired great deal of theoretical activity but also multiple experimental efforts to search for processes
that violate baryon number. 
Two classes of B-violating processes that are under active scrutiny are $\Delta B=1$ processes that involve proton decay~\cite{pati} and $\Delta B=2$ ones such as neutron-anti-neutron ($n-\bar{n}$) oscillation~\cite{marshak2, marshak3}. 
There is no experimental evidence at the moment for either. There is  plan for a very high sensitive search for $n-\bar{n}$ oscillation at the European Spallation Source (ESS) in Lund~\cite{milstead} 
whereas the Deep Underground Neutrino Experiment (DUNE)~\cite{DUNE} at Fermilab also plans to extend the sensitivity of the search
in Argon nuclei few times the current Super-Kamiokande bound~\cite{superK}, 
which in turn has considerably improved the old ILL bound~\cite{ILL}. 
There is also a recent search by using deuterium at SNO~\cite{SNO}. 
Clearly, the discovery of neutrino mass and the belief that neutrinos may be Majorana particles has provided another approach to the origin of matter~\cite{fuku}, leptogenesis which is a very active field. This uses the seesaw mechanism~\cite{seesaw1, seesaw2, seesaw3, seesaw4, seesaw5} which gives Majorana masses for neutrinos, which in some models lead to
observable neutron-anti-neutron oscillation.

Coming to $B$-violating processes, the BSM physics for  the two  classes of $B$-violation, proton decay and $n-\bar{n}$ oscillation, are very different and probe very different  microscopic distance scales in nature. 
While the simplest proton decay modes probe physics at the grand unified theory (GUT) scale of $10^{15}$ GeV 
(or distances of order $10^{-30}$ cm),  the $\Delta B=2$ processes such as $n-\bar{n}$ oscillation~\cite{marshak2, marshak3}
involve dimension 9 operators with six quarks and are suppressed by $M^5$  and probe physics around TeV to 100 TeV scale 
(or distance scales of order $10^{-18}-10^{-20}$ cm or so). 
This makes $n-\bar{n}$ transition of great interest since that would open up prospects for new kind of  BSM physics searches in colliders as well as other non-accelarator experiments. 
Neutrino mass on the other hand can probe a variety of scales from TeV to $10^{14}$ GeV.

In this note we focus on the question of how we can understand matter-anti-matter asymmetry in models with observable $n-\bar{n}$ oscillation. It is particularly important since it appears that canonical proton decay modes such as $p\to e^++\pi^0$ predicted by minimal GUTs do not seem to have much of a connection to the origin of matter due to the fact that they conserve $B-L$ symmetry.
As far as $n-\bar{n}$ oscillation goes however, there is no such obstacle. One proposal for understanding the origin of matter for this case is the post-sphaleron baryogenesis mechanism(PSB)~\cite{PSB, BDFM} where a real scalar particle, usually the $B-L$ breaking Higgs field, decays to six quarks (possibly via some intermediate states)  and six anti-quarks and in combination with other interactions in the theory that include CP violation lead to nonzero baryon asymmetry. When the scalar field acquires a vacuum expectation value (VEV), 
it leads to observable $n-\bar{n}$ oscillation. 
Alternatively, the scalar field could be replaced by a color neutral Majorana fermion which couples to three quarks, leading to both baryogenesis~\cite{babu, dev} as well as $n-\bar{n}$ oscillation. These are theories where typically all scales are similar and are in the multi-TeV range.
Another possibility is to have one diquark scalar with GUT scale mass decay to produce baryon asymmetry~\cite{babumoh} with another diquark scalar with TeV mass. We will call these split scale models for $n-\bar{n}$ oscillation. Baryogenesis has been explored in these models; 
see for example ~\cite{babumoh}. 
 Here we explore an alternative but attractive scenario of the Affleck-Dine (AD) baryogenesis~\cite{AD, DRT} and discuss its implications for specific baryon and lepton number violating processes and in particular for $n-\bar{n}$ oscillation.

There have been many realizations of  AD baryogenesis in the literature (for a review, see \cite{EM, am}). 
An essential part of AD mechanism is the existence of a flat direction carrying  baryon number which after inflation  dynamically generates baryon excess as it evolves with the Hubble expansion from suitable initial conditions~\cite{DRT}. Typically, one considers supersymmetric theories where there are many flat directions~\cite{DRT} for baryon number carrying super-partners to implement the AD mechanism.
It has also recently been pointed out that one could use the flat directions in Nambu-Goldstone bosons in theories with spontaneously broken global symmetries to implement the AD mechanism~\cite{hari}. 

Two parts to the discussion of AD baryogenesis are: the implementation of inflation followed by an epoch
where an AD field carrying the baryon number oscillates to generate baryon asymmetry. 
This eventually gets transmitted to the asymmetry of the SM baryons by the AD field decay via its coupling with SM fermions. 
In most models, the inflaton field and the AD field are different. 
However, there are models where the AD field and the inflaton field can be same \cite{Cline:2019fxx, Charng:2008ke, Hertzberg:2013jba, Takeda:2014eoa, Lin:2020lmr, stubbs, Kawasaki:2020xyf, Barrie:2021mwi}. 
In our discussion, we follow one such model proposed by Lloyd-Stubbs and McDonald~\cite{stubbs} with slight modification and apply it to various $B$ violating processes including $n-\bar{n}$ oscillation.
We note, however, that our discussion should also be applicable to other models of AD baryogenesis, although the detailed conclusions could be different.

%
To make the inflationary prediction to be consistent with the current Cosmic Microwave Background (CMB) observations, 
 we introduce nonminimal coupling of the AD field with gravity (see, for example, \cite{nobu1, nobu2}, references therein). 
We then try to connect the discussion to low energy baryon and lepton number violation. 
To explore the implications for various baryon and lepton number violating processes, we endow the AD field 
with the corresponding quantum number by coupling it to suitable operators. 
We then describe how baryon or lepton number is spontaneously broken to generate the observed asymmetry. 
 Several questions arise in such models.
For example, what is the scale of the violation of the quantum numbers ($B$ or $L$) compatible with constraints of adequate baryogenesis which can then determine whether the process can be observable in current searches. 
Secondly, after baryon or lepton number is spontaneously broken, there are $B$ or $L$ violating processes in the early universe
down to the decoupling temperature $T_D$ of those processes.
Since in AD baryogenesis gets transmitted to SM fermions at the reheat temperature $T_R$, 
one must have $T_D \ll T_R$ for the generated baryon or lepton asymmetry not to get erased. 
The reheat temperature $T_R$ is predetermined in a model from independent considerations. 
So it needs to be checked if in a given model there is washout of baryogenesis or not. 
We address these questions for several $B$ and $L$ violating examples 
by examining whether AD baryogenesis works or does not work, 
while yielding adequate $n_B/s$ together with the particular process being observable in the current searches.
 
After commenting on implications for nucleon decay and neutrino mass, we focus on $n-\bar{n}$ oscillation and 
 show that the preferred scenario  for AD baryogenesis in this case has a split scalar spectrum for the latter, similar to the one that arises in nonsupersymmetric SO(10) model proposed  in \cite{babumoh} mentioned above.
 This therefore makes the scenario embeddable into an SO(10) GUT theory. 
 In this case, the scalar submultiplets of the {\bf 126} Higgs field can play the role of inflaton and the field that decays to generate baryon asymmetry. An added {\bf 16}-plet Higgs helps to generate a VEV for the {\bf 126} field. We also discuss a $100$ TeV scale scenario where we find a narrow range where viable AD baryogenesis is compatible with observable $n-\bar{n}$ transition.

This paper is organized as follows: in Sec.~2, we describe the model used to discuss baryogenesis; 
in Sec.~3, we discuss the evolution of the universe in our model. 
In Sec.~4 we discuss how the Affleck-Dine baryognesis takes place in general class of such models. 
In Sec.~5, we discuss how the AD field acquires a VEV so that it gives rise to $B$ and $L$-violating processes at low energy. 
In Sec.~6, we discuss the various $B$ and $L$-violating processes and point out in Sec.~7 that a split scale GUT embeddable model provides the best opportunity for obtaining the adequate baryogenesis as in  the model of Ref.~\cite{babumoh} which provides a GUT-setting for our scenario and also discuss the viable 100 TeV scale scenario. We elaborate a bit more on the So(10) model in sec.~8 and 
Sec.~9 is devoted to a new $\Delta B=4$ process induced in the AD scenario we pursue and then we conclude our discussion in Sec.~10.

\section{2. The Model }  
While there are different ways to implement AD baryognesis, the model  presented here is a generalization of the work in~\cite{stubbs} which uses scalar field  $\Phi$ with the appropriate $B$ or $L$ quantum number, both as the inflaton and the AD field. 
We, non-minimally, couple the AD field to gravity so that it is consistent with CMB observations. 
Let us start by reviewing the results of Ref.~\cite{stubbs}. 
The starting Lagrangian for  $\Phi$ in this case is given by: 
\begin{eqnarray}
{\cal S}~=~\int d^4x\sqrt{-g}\left[-\frac{1}{2} M^2_P\, f \, R+\partial _\mu\Phi^\dagger\partial^\mu\Phi-V(\Phi)\right],
\end{eqnarray}
where $M_P=2.44 \times 10^{18}$ GeV is the reduced Planck mass, 
$f =1+2 \xi \frac{\Phi^\dagger\Phi}{M_P^2}$ with $\xi$ being non-minimal coupling to gravity.
We choose $V(\Phi)$  as in \cite{stubbs} 
\begin{eqnarray}
V(\Phi)~=~m^2_\Phi \Phi^\dagger\Phi - A(\Phi^2+\Phi^{\dagger 2}) +\lambda(\Phi^\dagger\Phi)^2. 
\end{eqnarray}

To discuss inflation in the model, we make transformation of the fields to go to the Einstein frame by  
$g^E_{\mu\nu}=g_{\mu\nu}/f$, 
which then leads to the following action $S_E$ in the Einstein frame,
\begin{eqnarray}
S_E~=\int d^4x \left[-\frac{1}{2}M^2_P R_E + 
\left( \frac{1}{f}+\frac{12 \xi^2}{f^2}  \frac{\Phi^\dagger \Phi}{M_P^2}\right)\partial _\mu\Phi^\dagger \partial^\mu\Phi -V_E(\Phi)\right],
\end{eqnarray}
where 
\begin{eqnarray}
V_E(\Phi)~=~  \frac{V(\Phi)}{ \left(1+2 \xi \frac{\Phi^\dagger \Phi}{M^2_P} \right)^2}.
\end{eqnarray}
To study the inflation picture and the AD mechanism, we switch to radial parameterization of $\Phi=\frac{1}{\sqrt{2}}|\Phi| e^{i\theta}$. The $|\Phi|$ field is then the inflaton field.
It is now clear that for large values of the field $|\Phi| \gtrsim M_P/\sqrt{\xi}$ in the early stage of the universe, the potential flattens out and is a constant to a good approximation driving the exponetial expansion of the universe - the inflationary phase.  
The inflation is essentially controlled by one free parameter $\xi$. 
The fits to observations such as the spectral index $n_s$ as well as the tensor-to-scalar ratio $r$ for a fixed number of e-folds $N_e$ 
in such a  model have been carried out in ~\cite{nobu1, nobu2}. 
The initial value of the inflaton field $|\Phi|$ is appropriately chosen to fit observations. 
For example, one bench mark choice of parameters that fits data is $\xi\sim 1600$ and $\lambda\sim 10^{-3}$
so that one gets $n_s=0.968$ and $r=0.003$ for $N_e=60$, which are fully consistent with observations~\cite{nobu1}. 
The $|\Phi|_{int}\sim 0.23 M_P$ for inflaton value at horizon exit and $|\Phi|_{end}\sim 0.029M_P$ at the end of inflation. 
We choose $|\Phi|_{end}$ as the initial value for the inflaton field in AD baryogenesis. 
The initial value of the phase of the $\Phi$ field can be chosen at random and
we choose it to be  $\theta = {\cal O}(1)\neq \pi/2$.  
Note the large value of the $\xi$ above. 
Clearly it raises the question of unitarity violation above a certain mass scale. 
This question has been analyzed for generic non-minimally coupled inflaton in Refs.~\cite{Barvinsky:2009ii, Ferrara:2010yw, Bezrukov:2010jz, Joergensen:2014rya}
and it has been noted that there is no real issue:
since during inflation the inflaton value is around the Planck scale, 
we estimate the effective cutoff to satisfy the unitarity by expanding the inflaton
around its background value, so that the effective cutoff is found to be the Planck scale.  
The second point we want to emphasize is that the presence of the $A$ term breaks the global baryon number symmetry carried by the rest of the Lagrangian and plays a crucial role in the baryon asymmetry generation.
This is also required by Sakharov's conditions for baryogenesis. 
It splits the masses of the real and imaginary parts of the $\Phi$ field. 
We will see later (Eq.~(10), (11) and below) that indeed $n_B$ is proportional to $A$. 

\section{3. Evolution of the universe in our picture}
In this model, there are four stages of the evolution of the early universe: 

\begin{enumerate} 
\item  For $|\Phi| \gtrsim M_P/\sqrt{\xi}$ when the non-minimal coupling in the Einstein frame leads to a constant $V(\Phi)$, it drives inflation as just noted in the previous section.

\item In the second phase, the value of $|\Phi|$ is still large  but not large enough to make the non-minimal gravity coupling dominate; instead the dominant term driving the evolution of the $|\Phi|$ is the $\lambda |\Phi|^4$ term.  Since the field  $|\Phi|$ has rolled down the potential and its value has become less than $M_P/\sqrt{\xi}$ the effect of the non-minimal coupling becomes unimportant and inflation ends. 
At the beginning of this stage, the real and imaginary parts of the field are already different due to the CP-violating $A$ term in the potential.
This asymmetry leads eventually to the baryon asymmetry of the universe and is the key idea in AD baryogenesis.

\item The third stage is where the quadratic term in the potential dominates over the quartic term leading to an oscillatory behavior of $|\Phi|$ (see below) and the universe behaves like it is matter dominated. This approximation of transition of the potential from being quartic dominated to quadratic dominated is called the threshold approximation in ~\cite{stubbs}.

\item The fourth  stage is when the AD field decays and reheat takes place. The reheat temperature determined by the decay width of the AD field will be denoted by $T_R$ and will determine the amount of baryon asymmetry generated. 
The Big Bang Cosmology era begins after this.

\end{enumerate}

To calculate the baryon asymmetry of the universe, one can make the so-called threshold approximation as has been done in ~\cite{stubbs}  and then one can  solve the time evolution  equations for the real and imaginary parts of the $\Phi$ field 
i.e.~$\phi_{1,2}/\sqrt{2} \equiv {\rm Re}[\Phi], {\rm Im}[\Phi]$.
We have solved these time evolution equations numerically to calculate the baryon asymmetry and we are in broad agreement
with the conclusions of Ref.~\cite{stubbs}. 
We first summarize the basic contents of the analytic solutions in the threshold approximation~\cite{stubbs}.
For $\phi_{1,2} \gtrsim \phi_* \equiv m_\Phi/\sqrt{\lambda}$, the quartic term in the potential dominates. 
When that happens, one can see as follows that $\phi_{1,2}$ decrease with the expansion of the universe as $\phi_{1,2}\propto 1/a$.  
To see this, note that 
\begin{eqnarray}
\dot{\rho}+3H(\rho+p)=0
\end{eqnarray}
where $\rho= \dot{\Phi}^\dagger \dot{\Phi} +V$ and $p= \dot{\Phi}^\dagger \dot{\Phi}-V$ are the energy density
and pressure of the universe at early times. 
It is known that when the quartic term dominates the potential during the inflaton oscillation,
the equation of state behaves like the radiation dominated era, $p=\rho$/3, leading to
\begin{eqnarray}
\dot{\rho}+4H\rho=0
\end{eqnarray} 
and  $\dot{\Phi}^\dagger \dot{\Phi} \sim 2 V$.
This gives $\rho a^4 \sim 3 V a^4 \sim 3 \lambda (\Phi^\dagger \Phi)^2 a^4 ={\rm constant}$, 
which implies $|\Phi|\propto 1/a$.
Since $ |\Phi| \propto 1/a$ as the universe expands, the $|\Phi|$ value  goes down and at some point 
for $\sqrt{2}|\Phi|= \phi_* = m_\Phi/\sqrt{\lambda}$ and the quadratic term starts dominating the potential. 
The field amplitudes at $a_*$, which is the expansion rate when $\sqrt{2}|\Phi|= \phi_* $, are expressed as 
\begin{eqnarray} 
  \phi_{i,*}=\left(\frac{a_I}{a_*} \right) \phi_{i, I} = \left(\frac{\phi_*}{\phi_I} \right) \phi_{i, I},
\end{eqnarray}
where $\phi_{i,I}$ is the initial values of $\phi_i$ at $a=a_I$, and $\phi_I=\sqrt{(\phi_{1,I})^2+(\phi_{2,I})^2}$.
To follow the evolution of $\phi_{1,2}$ after this point $\phi_*$, we use the quadratic term to solve the evolution equation
as will be done in the next section. 
%
%

\section{4. Evolution of AD field after stage 2 and  baryogenesis} 
To study baryogenesis, we look at the time evolution of the real and imaginary parts of the field $\Phi$ 
by using the following equations of motion,
\begin{eqnarray}
\ddot{\phi_1}+3H\dot{\phi_1}=-m^2_1\phi_1-\lambda (\phi^2_1+\phi^2_2)\phi_1,\\\nonumber
\ddot{\phi_2}+3H\dot{\phi_2}=-m^2_2\phi_2-\lambda (\phi^2_1+\phi^2_2)\phi_2,\\\nonumber
\end{eqnarray}
where $m^2_1=m^2_\Phi-2A$ and $m^2_2=m^2_\Phi +2A$.
We also follow this evolution numerically.  
To get an analytical solution, we can  neglect the quartic terms since as argued above at this stage the contribution of the quartic term is very small compared to the quadratic term.  
Then for 
 $H \ll m_\Phi$,  
one can write approximate solutions for  $\phi_{1,2}$ components of the fields to be:
\begin{eqnarray}\label{phi12}
\phi_{i}(t)\simeq \phi_{i,*} \left(\frac{a_*}{a}\right)^{3/2}~{\cos}(m_{i}(t-t_*))
= \phi_{i, I} \left(\frac{\phi_I}{\phi_*} \right)^{1/2} \left( \frac{a_I}{a}\right)^{3/2} ~{\cos}(m_{i}(t-t_*)). 
\end{eqnarray}
Note the difference between the evolution equations for the real and imaginary parts of $\Phi$. 
Because of this difference (and the initial value of $\theta={\cal O}(1)\neq \pi/2$), 
nonzero baryon number of the universe will be generated. 
In what follows, we parameterize $A =\epsilon M^2_\Phi$  with $0 < \epsilon \ll 1$. 
Baryon number  asymmetry is given by ${n_B}(t)= Q_\Phi (\dot{\phi_1}\phi_2-\dot{\phi_2}\phi_1)$. 
We can then rewrite the time evolution of $n_B(t)$ using the above equations of motion as
\begin{eqnarray}
\dot{n}_B+3Hn_B &= &2Q_\Phi  {\rm Im}\left(\frac{\partial V}{\partial \Phi^\dagger} \Phi^\dagger \right)
=4 Q_\Phi \, A \, \phi_1(t) \, \phi_2(t) \nonumber\\
&\simeq& 4 Q_\Phi \, A \, \phi_{1,I} \, \phi_{2,I} \left(\frac{\phi_I}{\phi_*} \right) \left( \frac{a_I}{a(t)}\right)^3 
~{\cos}(m_{1}(t-t_*)) ~{\cos}(m_{2}(t-t_*)).
\end{eqnarray}
The baryon asymmetry is generated for $t > t_*$. 
Defining the co-moving asymmetry $N_B= \left(\frac{a(t)}{a_I}\right)^3n_B(t)$, 
we evaluate the baryon asymmetry by 
\begin{eqnarray}
N_B (t)&\simeq& 
2Q_\Phi \int_{t_*}^t dt' \left(\frac{a(t')}{a_I}\right)^3 {\rm Im}\left(\frac{\partial V}{\partial \Phi^\dagger} \Phi^\dagger \right)
 e^{- \Gamma_\Phi (t'-t_*)} \nonumber \\
&\simeq &
4 Q_\Phi \, A \, \phi_{1,I} \, \phi_{2,I} \left(\frac{\phi_I}{\phi_*} \right) 
 \int_{t_*}^t dt' \, {\cos}(m_{1}(t'-t_*)) \, {\cos}(m_{2}(t'-t_*))  \, e^{- \Gamma_\Phi (t'-t_*)} , 
\end{eqnarray}
where we have introduced the decay factor $e^{- \Gamma_\Phi (t'-t_*)}$ 
since the inflaton decays to the SM particles with its decay width $\Gamma_\Phi$ and 
its amplitude exponentially damps for $t > 1/\Gamma_\Phi$. 
In fact, we have a simple expression of  the time-integral for $t > 1/\Gamma_\Phi$\footnote{
This analytic expression is our new finding, which allows us to evaluate the resultant baryon asymmetry
for any choice of $\epsilon, \gamma \ll1$.
}: 
\begin{eqnarray}
{\cal I}&\equiv&\int_{t_*}^t dt' \, {\cos}(m_{1}(t'-t_*)) \, {\cos}(m_{2}(t'-t_*))  \, e^{- \Gamma_\Phi (t'-t_*)} \nonumber\\
&\simeq & \frac{\gamma}{2 m_\Phi}
\left(
 \frac{1}{2+\gamma^2-2 \sqrt{1-4 \epsilon^2}} + \frac{1}{2+\gamma^2+2 \sqrt{1-4 \epsilon^2}} 
\right), 
\end{eqnarray}
where $\gamma \equiv \Gamma_\Phi/m_\Phi \ll 1$ for a narrow decay width. 
We can see that for $2 \epsilon \gg \gamma$ (or, equivalently, $2 A \gg \Gamma_\Phi m_\Phi$), 
${\cal I} \simeq \frac{\gamma}{8 \epsilon^2 m_\Phi}$ 
while ${\cal I} \simeq \frac{1}{2 \gamma m_\Phi}$ for $2 \epsilon \ll \gamma$. 
Since $N_B$ is proportional to $A= \epsilon m_\Phi^2$ with $\epsilon \ll1$, 
we consider the case of $2 \epsilon \gg \gamma$ to obtain the resultant baryon asymmetry as much as possible.

The total baryon asymmetry transferred to the SM thermal plasma at the time of reheating is given by
\begin{eqnarray}
   n_B = N_B \left(\frac{a_I}{a_{R}} \right)^3=N_B \left(\frac{a_I}{a_*} \right)^3 \left(\frac{a_*}{a_{R}} \right)^3
   \simeq N_B \left(\frac{\phi_*}{\phi_I} \right)^3 \left(\frac{H_{R}}{H_*} \right)^2, 
\end{eqnarray}
where we have used $a \propto t^{2/3} \propto H^{-2/3}$ for the inflaton oscillations of Eq.~(9). 
Using the Friedmann equation, we have $H_{R}^2= \frac{\pi^2}{90} g_* \frac{T_{R}^4}{M_P^2}$ 
with $g_* \simeq 100$ is the relativistic degrees of freedom of the SM thermal plasma 
and $H_*^2 \simeq \frac{m_\Phi^2 \phi_*^2}{6 M_P^2}$. 
We now obtain the final expression for $n_B/s$ with the entropy density of the SM thermal plasma, 
$s=\frac{2 \pi^2}{45} g_* T_{R}^3$, to be
\begin{eqnarray}
   \frac{n_B}{s} \simeq \frac{3}{8} \sqrt{\frac{\pi^2}{90} g_*} \frac{Q_\Phi}{\epsilon}
   \frac{T_{R}^3}{m_\Phi^2 M_P} \sin(2 \theta) 
  \simeq 10^{-13}\, \frac{Q_\Phi}{\epsilon}\, \left(\frac{T_R}{10^{12}~{\rm GeV}}\right)^3
  \left(\frac{10^{15}~{\rm GeV}}{m_\Phi}\right)^2. 
  \label{nB}
\end{eqnarray}
For $\epsilon = 10^{-3}$ and  $\sin(2 \theta) \sim 1$\footnote{
A very small initial $\theta$ may generate iso-curvature fluctuation which is too large
to be consistent with the CMB observations \cite{Barrie:2021mwi}. 
}, 
this gives the right order of magnitude for $n_B/s \simeq 10^{-10}$.  
We emphasize that we cannot make $\epsilon$ too small since in the limit of $\epsilon=0$, 
the baryon asymmetry vanishes (see Eqs.~(11) and (12)). 
We will use this value for $m_\Phi$ motivated by GUT theories, 
although we will give some examples where $m_\Phi$ is much lower. 
In what follows we will take $\epsilon$ accordingly but choose the actual magnitude to make $n_B/s$ to fit observations as well as to make $B$ and $L$ violating process in question observable compatible with above constraints on it.

%




\section{5. Origin of VEV $\langle \Phi \rangle$}
The next question is how to generate a VEV for the $\Phi$ field.
There are two ways to accomplish that: (i) first way is to choose the mass term in Eq.~(2) to be negative and rerun the $\phi_{1,2}$ evolutions again; (ii) a second way is to couple $\Phi$ to a new field ($\chi$) with $B=-1$
and give $\chi$ a VEV which will then induce a type of VEV, 
$\langle \Phi \rangle = \frac{\tilde {m}v^2_\chi}{m^2_\Phi}$, as we will see below. 
The latter case has the advantage that it does not affect the evolution of the $\phi_{1,2}$ fields
since the $\chi$ field decouples and leaves only an inconsequential $\Phi$ tadpole at lower energies.  
To see this in detail, 
we add the following potential to Eq.~(2): 
\begin{eqnarray}
V(\Phi, \chi)~=~-\tilde{m}\chi\chi\Phi+h.c.+\lambda_\chi (|\chi|^2-v^2_{\chi})^2
\end{eqnarray}
After integrating the $\chi$ field, we obtain tadpole terms, $\tilde{m} v_\chi^2 \Phi +h.c.$,
which leads to $\langle \Phi \rangle =\frac{\tilde{m}v^2_\chi}{m^2_\Phi}$.  

Let us show that adding a linear term after the $\chi$ field is integrated out, 
the evolution of the AD field is still dominated by the $\Phi^4$ and $\Phi^2$ terms as before. 
Thus, our analysis in the previous section remains the same. 
To show that, let us write the potential in the presence of the linear term and set $\phi_2=0$, for simplicity.
\begin{eqnarray}
V(\phi_1)~=~ -M^3\phi_1+\frac{1}{2} m^2_1 \phi^2_1+ \frac{1}{4} \lambda \phi_1^4, 
\end{eqnarray}
where $M^3=\sqrt{2} \tilde{m} v_\chi^2$. 
We assume $M \sim m_1$ and parametrize $\lambda =\left(\frac{m_1}{M} \right)^6 \delta \sim \delta \ll1$. 
Solving the stationary condition, we find $\langle \phi_1 \rangle=\frac{M^3}{m_1^2}(1-\delta + 3 \delta^2)$ 
up to ${\cal O}(\delta^3)$. 
Expanding the field $\phi_1$ around its VEV, $\phi_1= \frac{M^3}{m_1^2}(1-\delta + 3 \delta^2 + \varphi)$, 
we express the potential as 
\begin{eqnarray}
V(\phi_1)~ \simeq~\frac{M^6}{m_1^2} 
\left(
-\frac{1}{2} + \frac{1}{2} \varphi^2 + (\delta-\delta^2) \varphi^3 + \frac{1}{4}\, \delta \, \varphi^4
\right). 
\end{eqnarray}
Note that the coefficient of $\varphi^3$ is the same order of that of $\varphi^4$. 
Hence, the $\varphi^3$ term dominates over the $\varphi^4$ term for $\varphi <1$. 
However, in this case, the potential is dominated by $\varphi^2$ term. 
For $\varphi > 1/\sqrt{\delta}$, the $\varphi^4$ term dominates over the $\varphi^2$ term. 
Therefore, the $\varphi^3$ term can never dominates the potential.

\section{6. Connecting to $B$ and $L$-violating processes} 
To study the  phenomenological implications of the implementation of AD mechanism this way, we endow the $\Phi$ field 
with appropriate $B$ and/or $L$  charges and study the effect of the baryogenesis constraints on the magnitude
of the relevant process e.g.~whether it is observable and derive conclusions about whether AD baryogenesis is viable for a particular model. 
For this analysis, we start with the $\Phi$ coupling to the SM (or slightly beyond SM) fields, given by $\Phi {\cal O}_d/\Lambda^{d-3}$, 
where ${\cal O}_d$ is the $B$ or $L$-violating operator with a mass dimension $d$. 
We will then use the three constraints to see if AD baryogenesis for a particular operator leads to observable $B$ or $L$-violation. 
The three constraints are:
\begin{itemize}

\item Adequate amount of baryon asymmetry i.e. $\frac{n_B}{s}\simeq 10^{-10}$ using the formula of  Eq.~(\ref{nB});

\item The baryon asymmetry generated by the AD field should not be washed out 
when $\langle \Phi \rangle = v_\Phi\neq 0$  since this VEV leads to processes in the early universe that violate baryon or lepton number;

\item  The $B$ or $L$-violating  process generated by $\langle \Phi \rangle \neq 0$ should be in the observable range of current or planned experiments.

\end{itemize}

The expression for $n_B/s$ is already given in Eq.~(\ref{nB}) above. 
Since below a certain temperature, the model has $B$ violating interactions, they can in principle erase the generated baryon asymmetry 
via the so-called washout processes if they are in equilibrium.  
To avoid the washout, the decoupling temperature for the relevant $B$ or $L$-violating process must be above the reheat temperature $T_R$ 
since the baryon asymmetry generated by the AD (inflaton) field is transmitted to the the SM sector by the reheating. 
In this discussion, we will assume  for definiteness that $m_\Phi\sim v_\Phi \sim 10^{15}$ GeV except one example below and choose the parameter $\epsilon \leq 0.001-0.1$. 
Once we choose $m_\Phi$ and $\epsilon$, this leaves us with a free parameter $\Lambda$ which will have a lower limit to satisfy the $T_R$ constraint determined from $n_B/s$. 
Since this is the scale of the higher dimensional operator coupling to $\Phi$, it together with $v_\Phi$ determines
whether the process in question is observable or not in current searches. 
Below we give three examples of processes where we apply this strategy and  find the values of $\Lambda$ using $T_R$ whose upper bound is determined from the $n_B/s$ formula and $\epsilon \leq 0.001-0.1$. 
As we show in the examples below, the higher the dimension of the operator ${\cal O}_d$, 
the lower the probability of it being observed. 
We then follow it up with an example that uses lower values of $m_\Phi$ and $v_\Phi$ (in the 100 TeV range)
to see whether such a scenario works.

To carry out this program, the first point which is common to all scenario is the value of $T_R$ from the $n_B/s$ expression. There is some small difference in the values of $\Lambda$ for our choice of benchmark point of $m_\Phi$ and $v_\Phi$. 
Using $n_B/s\sim 10^{-10}$, we find that $T_R \leq 10^{12}$ GeV for $\epsilon \leq 0.001$. 
Once we are given the reheat temperature $T_R$, we can use the formula $T_R\simeq \sqrt{\Gamma_\Phi M_P}$ to estimate the scale of $\Lambda$.
Typically, it turns out to be of order or greater than $10^{15}$ GeV for this choice of parameters $m_\Phi$ and $\epsilon$ depending on the dimensionality of the operator.
This then makes it clear why the higher the dimension of the operator coupled to $\Phi$ is, 
the $B$ or $L$-violating processes generated are suppressed. 
On the other hand in split scalar models, the dimension of the operator ${\cal O}_d$ is lower in the early universe
and the physical process gets enhanced observability since at zero temperature, the effective scale is then of the form, 
$\Lambda^n M^m$, where $M$ is a mass scale of new particles, which are at a lower scale. The latter can be chosen in the TeV range. 
Sometimes the process can get so enhanced that AD mechanism for baryogenesis is not viable in that case. 
We illustrate this in the following examples.

\subsection{6a. Nucleon decay via $B-L$ violating operators}  We do not consider the canonical proton decay operators of type $QQQL$, etc.~that lead to $p\to e^++\pi^0$, since it conserves $B-L$ and any asymmetry generated by this will be washed out by sphaleron effects.  
Instead a $d=7$ operator of the type is ${\cal O}_7 = Qe^c(d^cd^c)^*H^*$ breaks $B-L$ and does not suffer from this problem. 
This leads to a relevant $\Phi$ coupling as 
\begin{eqnarray}
{\cal L}_\Phi \sim \frac{1}{\Lambda^4}\Phi Qe^c(d^cd^c)^*H^* .
\label{QeddH}
\end{eqnarray}
We estimate the decay width of $\Phi$ as $\Gamma_\Phi \sim PS^{(5)} \frac{m_\Phi^9}{\Lambda^8}$, 
where 
\begin{eqnarray}
PS^{(n)} = \frac{1}{2 (4 \pi)^{2n-3} \Gamma(n) \Gamma(n-1)}
\end{eqnarray}
 is the phase space factor 
for $n$-body decay of $\Phi$ (see, for example, \cite{Bashir:2001ad}).
Using our argument above, we find that $\Lambda \geq 5.7 \times 10^{15}$ GeV for $T_R \leq 10^{12}$ TeV 
with $m_\Phi = 10^{15}$ GeV and $\epsilon=0.001$.  
This leads to an inaccessibly much too long a lifetime for the process $n\to e^-\pi^+$ induced by this operator after Higgs $H$ VEV and the $\Phi$ VEV are used. 
The lifetime is roughly predicted to be around $10^{41}$ years, clearly unobservable.

Now suppose we replace any pair of fermions in the operator (say $d^c d^c$) by a TeV scale scalar. 
Then the effective coupling for $B-L$ violating nucleon decay can be written as $\frac{1}{\Lambda^2}\Phi \Delta_{dd} Qe^cH^*$. 
Note the lower power of $\Lambda$. 
Since the decay width of $\Phi$ is roughly estimated as $\Gamma_\Phi \sim PS^{(4)}\frac{m_\Phi^5}{\Lambda^4}$, 
we find $\Lambda \geq 4.2 \times 10^{15}$ GeV by using $T_R \simeq \sqrt{\Gamma_\Phi M_P} \leq 10^{12}$ GeV. 
If a Yukawa coupling $\Delta_{dd} d^c d^c$ is introduced, after integrating $\Delta_{dd}$ out, 
we obtain the effective operator of Eq.~(\ref{QeddH}) with the replacement $\Lambda^4 \to \Lambda^2 m_\Delta^2$. 
After substituting $\langle \Phi \rangle =10^{15}$ GeV, $\langle H \rangle =174$ GeV,
$\Lambda  =10^{16}$ GeV and $m_\Delta=1$ TeV, we find the neutron lifetime to be around $10^{10}$ years,
which is too short and is ruled out by the current data. 
Thus  unless we artificially push the $\Lambda$ to Planck scale, this operator cannot help us to use AD baryogenesis. This is a new result with future implications i.e.~if $n\to e^-+\pi^+$ is discovered in ongoing nucleon decay searches, 
one cannot apply AD baryogenesis along with this operator in the simple way discussed.

\subsection{6b. Neutrino mass via $L$-violating operators} 
We may choose the $\Phi$ field to have $L=2$ with interaction  of two kinds: first using the Weinberg operator as follows:
\begin{eqnarray}
{\cal L}_\Phi~ = ~ \frac{1}{\Lambda^2}\Phi LHLH~+~h.c.
\end{eqnarray}
In this case, one can use the AD mechanism to generate a lepton asymmetry, 
which can be converted by the sphaleron process to baryon asymmetry. 
The reheating temperature is estimated by $T_R \simeq \sqrt{\Gamma_\Phi M_P}$ 
with the decay width $\Gamma_\Phi \sim PS^{(4)}\frac{m_\Phi^5}{\Lambda^4}$,
while the decoupling temperature of the washout process induced by the operator, $ \frac{v_\Phi}{\Lambda^2}LHLH$, 
is estimated as $T_D \sim \frac{\Lambda^4}{v_\Phi^2 M_P}$. 
Imposing the condition, $T_R < T_D$, we find $\Lambda > 1.9 \times 10^{15}$ GeV. 
%
Using the Weinberg operator this gives $m_\nu < 0.017$ eV, 
which is below that required for fitting the atmospheric neutrino oscillation data. 
Thus this way of doing leptogenesis to solve the baryon asymmetry problem is likely problematic. If we choose, smaller $\Lambda$ it will give too large a $T_R$ and hence an unacceptable value for $\epsilon$ for getting adequate $n_B/s$.

A second possibility for neutrino mass is to consider an operator that involves the right handed neutrino field $N$ of the form $\Phi NN$. 
In this case, we have $\Gamma_\Phi\simeq \frac{1}{12\pi} m_\Phi$ (assuming $m_\Phi \geq 2 M_N$), 
so that to satisfy the condition $2 A=2 \epsilon m_\Phi^2 \gg m_\Phi \Gamma_\Phi$ and $\epsilon \ll1$,
we can only have $\epsilon \sim 0.1$. 
Using $T_R=\sqrt{\Gamma_\Phi M_P} \simeq \sqrt{\frac{m_\Phi M_P}{12 \pi}}$ and Eq.~(\ref{nB}) with $\epsilon \sim0.1$, 
we find no solution for $m_\Phi < M_P$ to get $n_B/s\sim 10^{-10}$. 
%

\section{7.  Neutron-anti-neutron oscillation} 
To explore the implications for $n-\bar{n}$ oscillation, we will assume that the $\Phi$ field has $B=2$
and couples to $B=-2$ operators.  
These operators depend on the kind of scalar spectrum and can be of  the following three types in extensions of the SM.
Their strength will depend on scalar masses and their cosmological impact will depend on the temperature of the universe. 
We will look at the constraints on the parameters charactering the operator i.e.~$m_\Phi$, the mass of the AD field; 
$\Lambda$ the scale of $B-L$ violation, and $v_\Phi$ the VEV of the $\Phi$ field. 
Let us list the following three generic  scenarios. 
In each case, the gauge invariant operator involving the $\Phi$ field and other relevant fields will be as follows. 
As before,  from these operators $n-\bar{n}$ oscillation will arise 
when $\Phi$ field acquires a VEV thereby breaking baryon number by two units.

\begin{description}

\item[(i)] 
The first operator  is a six quark operator i.e.~$\frac{1}{\Lambda^6}\Phi uddudd$ 
where $u,d$ are the right handed parts of the SM quark fields and $\Lambda$ is the scale of new physics 
that gives rise to $\Delta B=2$ forces. 
This can for example be the case, when all the diquark Higgs fields connecting to two quarks (e.g.~$uu, ud, dd$) 
have same masses  of order of or lower than the GUT scale. 
We will take this operator involving only right handed quark fields for demonstrating AD baryogenesis in our model, 
even though several other kinds of operators are also possible~\cite{shrock}.  
Our conclusions will not depend on these details as we show below.

\item[(ii)] 
The next class of operator is a four quark plus a single diquark Higgs e.g. $\frac{1}{\Lambda^4}\Phi \Delta_{dd} uudd$, where $\Delta_{dd}$ decays to two right-handed down quarks. If $\Delta_{dd}$ mass is at the TeV scale, it will generate an $n-\bar{n}$ transition after $\Phi$ acquires a VEV.

\item[(iii)] 
Finally we have the lowest dimensional operator involving two diquark scalars plus two quarks e.g.~$\frac{1}{\Lambda^2}\Phi\Delta_{ud}\Delta_{ud} dd$, which again will lead to an $n-\bar{n}$ process in same way as the above cases. 
This is similar to a split scale mechanism for $n-\bar{n}$ oscillation proposed in the context of an SO(10) model in Ref.~\cite{babumoh}.

\end{description}

To make the discussion tractable, we consider two cases for $\Lambda$, which is the mass scale
for $n-\bar{n}$ oscillation with baryon asymmetry generated by the AD mechanism. 
For the first case, we consider $m_\Phi \simeq 10^{15}$ GeV and $v_\Phi\simeq 10^{15}$ GeV as before 
since we would like to understand the scale as having originated from the grand unification. 
For the second case, we leave $m_\Phi$ as a free parameter and consider the scale  $\Lambda$ being lower e.g.~$10^5$ GeV
since this is an example  of  the class of models (see Ref.~\cite{marshak3}) which have been considered widely 
in the field over the years.

We find that (as explained below) the $n-\bar{n}$ transition amplitude is not observable for both scenarios (i) and (ii) above because of the same argument as in Sec.~6 for $B-L=2$ nucleon decay. 
To repeat the argument, to get $n_B/s$ right without choosing too low an $\epsilon$, we have $T_R \lesssim 10^{12}$ or lower 
for $\epsilon \leq 0.001$. 
This implies that $\Lambda \gtrsim 10^{14.5}$ GeV and hence $n-\bar{n}$ transition is highly suppressed. 
The numbers for case (ii) operator are somewhat different but in the end the process is suppressed 
due to a high $\Lambda \sim 10^{15}$GeV and an effective 
$G_{\Delta B=2} \sim 10^{-51}$ GeV$^{-5}$. 
This therefore is not interesting for us.
%

However in case (iii), if we choose $T_R\simeq 10^{12}$ GeV for $\Gamma_\Phi \sim PS^{(4)}\frac{m_\Phi^5}{\Lambda^4}$, then we get $\Lambda\simeq 4.2 \times 10^{15}$ GeV. 
Due to split scale scenario, for this process, the strength of $n-\bar{n}$ comes out to be $G_{\Delta B=2}\simeq \frac{v_\Phi}{\Lambda^2 M^4_{ud}}\sim  10^{-28}$ GeV$^{-5}$ which brings it to the observable range if we keep the $\Delta_{ud}$ masses in the TeV range, as has been shown in Ref.~\cite{babumoh}.

\subsection{7a Case (i) with all scales in the $\sim 100 $ TeV range} 
To see if observable  $n-\bar{n}$ oscillation is compatible with viable AD baryogenesis in this case, we keep $m_\Phi, v_\Phi $ and $\Lambda$ in the range of $\sim 100$ TeV and impose the no-washout condition i.e. $T_R \ll T_D$, 
so that when the $\Delta B=2$ processes involving quarks appear, their strength has become so weak that they never get into equilibrium to erase the AD generated baryon asymmetry.  
(For a detailed discussion of wash-out in the case of general $n-\bar{n}$
 theories, see~\cite{hati}).
The reheating temperature is estimated by the decay width of $\Phi$ to 6 quarks as 
$ \Gamma_\Phi \sim PS^{(6)} \frac{m_\Phi^{13}}{\Lambda^{12}}$.
We estimate the decoupling temperature of the baryon number violating processes such as $qq \to q^c q^c q^c q^c$ by
\begin{eqnarray}
     T_D^3 \, \langle \sigma v_{rel} \rangle \sim H(T_D), 
\end{eqnarray}
where $\langle \sigma v_{rel} \rangle \sim  PS^{(4)} \frac{v_\Phi^2}{\Lambda^{12}} T_D^8$ 
is the thermal-averaged cross section times relative velocity of the process $qq \to q^c q^c q^c q^c$. 
For the following ranges of the parameters,
\begin{eqnarray}
     100 \leq \Lambda[{\rm TeV}] \leq 1000, \; 
     100 \, {\rm GeV} \leq m_\Phi \leq \Lambda, \; 
     100 \, {\rm GeV} \leq v_\Phi  \leq 1000 \, {\rm TeV},   
\end{eqnarray}
we have performed the random parameter scan to select the parameter set 
which satisfies $T_D > 10 \, T_R$ and $n_B/s =10^{-10}$ with $10^{-4} \leq \epsilon \leq 0.1$.\footnote{
Here, we have imposed $m_\Phi < \Lambda$ from the theoretical consistency for our effective operator analysis. 
We find that the resultant region for $m_\Phi$ to satisfy the conditions is limited to be 
in the range of $0.63 \lesssim \frac{m_\Phi}{\Lambda} \lesssim 0.85$.
}
Using the resultant parameter set, we estimate the $n-\bar{n}$ oscillation time by
\begin{eqnarray}
    \tau_{n-\bar{n}}^{-1} \simeq   G_{\Delta B=2} \, |{\cal M}|, 
 \end{eqnarray}
where $G_{\Delta B=2} = \frac{v_\Phi}{\Lambda^6}$, and ${\cal M}$ is the neutron-anti-neutron transition matrix elements
for which we employ a Lattice QCD calculation result, ${\cal M} = -\frac{26}{4} \times 10^{-5}$ GeV$^6$ \cite{Rinaldi:2019thf}. 
Fig.~1 shows the parameter scan result for the $n-\bar{n}$ oscillation time. 
Given the constraints, we find a narrow range of parameters where $n-\bar{n}$ oscillation is observable.

 \begin{figure}[!h]
  \centering
 \includegraphics[width=0.7\linewidth]{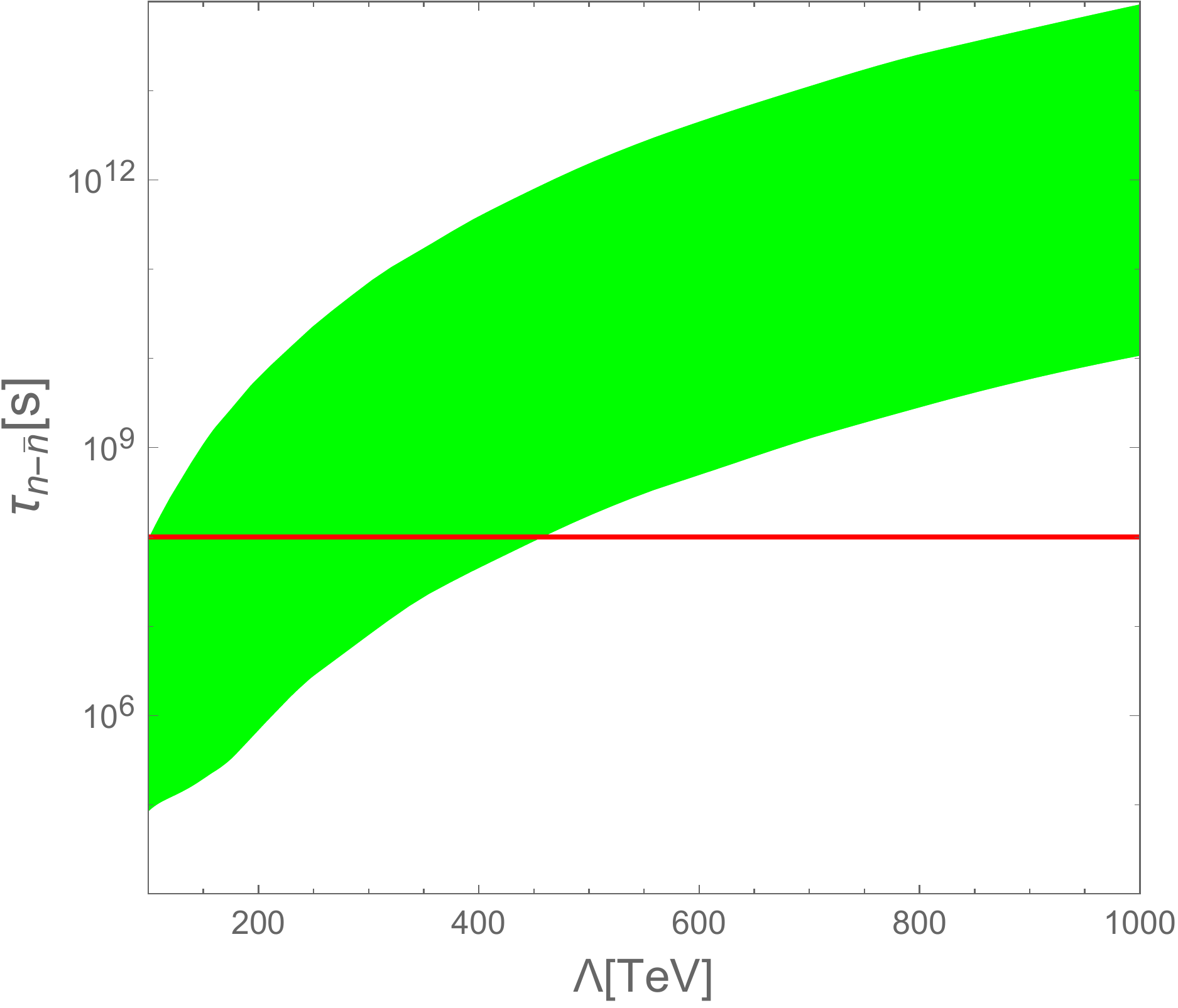}
  \caption{$n-\bar{n}$ oscillation time for the model parameters around the 100 TeV range. The region above the red line are allowed by current experiments. The ESS experiment will probe a significant part of the unexplored  oscillation time range.} 
  \label{tau}
  \end{figure}

\subsection{7b. Split scale scenario} 
In this case, the reheat temperature can be calculated once we have the $\Phi\to d^cd^c \Delta_{ud}^* \Delta_{ud}^*$ decay width is known. Here the $\Delta_{ud}$ coupling is given by $\Delta_{ud}u^cd^c$. 
The reheat temperature  can now be obtained from
\begin{eqnarray}
\Gamma_\Phi ~\sim ~PS^{(4)}\frac{m_\Phi^5}{M^4_{\Delta_{dd}}}.
\end{eqnarray} 
Setting all dimensionless couplings to one and $M_{\Delta_{dd}}\sim M_U =3 \times 10^{15}$ GeV,  
which is the gauge coupling unification scale found in Ref.~\cite{babumoh}, we find
$T_R \sim 2.0 \times 10^{12}$ GeV. 
%
%
We then compute the dominant washout process $\Delta_{ud}+d^c\to \Delta_{ud}^*+\bar{d^c}$ and it is easy to see that by the above reheat temperature this process is safely out of equilibrium. The actual temperature at which this process goes out of equilibrium is 
found to be $T_D \sim \frac{M^4_{\Delta_{dd}}}{v_\Phi^2 M_P} \simeq 3.3 \times 10^{13}$ GeV, 
which is above the reheat temperature $T_R$ for this case. 
Thus the split scale scenario of Ref.~\cite{babumoh} does provide a viable AD mechanism for baryogenesis.

\section{8.  Embedding of AD with split scale for $n-\bar{n}$  in SO(10)} 
We find it interesting that the split scalar spectrum that works best for AD baryogenesis with $n-\bar{n}$ oscillation   in this paper is similar to the SO(10) model in~\cite{babumoh} where the split scalar spectrum gives observable $ n-\bar{n}$ oscillation. We consider as in \cite{babumoh} a model with {\bf 10}, {\bf 126} and {\bf 54} Higgs fields. 
We add a second ${\bf 126^\prime}$ to the model to generate  the $A$ term. 
We also add a ${\bf 16}$ Higgs to generate the tadpole term for the ${\bf 126}$ Higgs field. 
It is well known that in this model  both {\bf 10} and {\bf 126} only couple to the SM fermions in {\bf 16} spinor of the SO(10). The {\bf 126} under the $SU(2)_L\times SU(2)_R\times SU(4)_c$ group consists of the multiplet $({\bf 1}, {\bf 3},{\bf 10})$ which has submultiplets which we denote by $\Delta_{\nu^c\nu^c}$, $\Delta_{u^cd^c}$ $\Delta_{u^cu^c}$ and $\Delta_{d^cd^c}$. We identify the  $\Delta_{\nu^c\nu^c}$ field as the inflaton and the AD field $\Phi$. The relevant terms  in the Higgs potential in Eq.(2) are easy to discern in the SO(10) Higgs potential. The $A$ term which breaks baryon number can arise from a coupling of type ${\bf 126}^2{\bf \bar{126^\prime}}^2$ term after two of the $\Delta^\prime_{\nu^c\nu^c}$ terms acquires a VEV. 
We can get the rest of the coupling terms in the potential small so that we can follow the evolution of the $\Phi$ field discussed above. 

It was shown in \cite{babumoh} that this model leads to coupling unification with $\Delta_{u^cd^c}$ field at the TeV scale along with a complex weak triplet scalar $\Delta$ ({\bf 1},{\bf 3},0) and a second Higgs doublet $H$ ({\bf 1},{\bf 2},1/2) included at this scale. The SO(10) gauge symmetry breaks down to the standard model below the GUT scale.

Processes such as $\Delta_{ud}+d^c\to \Delta_{ud}^* + \bar{d}^c$ go out of equilibrium below $T \simeq M_{\Delta_{dd}}$ 
for $M_{\Delta_{dd}} \sim 10^{15}$ GeV. 
This is far above the reheat temperature of about $10^{12}$ GeV or so to cause any washout.
So these $B$-violating processes  do not affect the AD generated baryons.


\section{9. $\Delta B=4$ process} Note that due to the presence of the $A$ term in the AD Higgs potential  has $B=4$, it can mediate a  super-super-weak process  such as $n+n\to \bar{n}+\bar{n}$ decay of nuclei. 
Since a baryon violating term in the potential involving the AD field is a generic feature of this model, we expect processes with $\Delta B= 4,6,...$.

 In our particular example, the strength of this process is given by
\begin{eqnarray}
G_{nn\to \bar{n}\bar{n}}\sim\left(G^2_{nn\Phi} \frac {\epsilon}{m^2_\Phi}\right)
\end{eqnarray}
where $G_{nn\Phi}$ is the dimensionless coupling which represents the $\Phi$ coupling to dressed six quark operator that gives rise to $n-\bar{n}$ oscillation. Using typical nuclear binding energy as the nuclear energy scale, we then estimate an order of magnitude for the nuclear disintegration rate via the $nn\to \bar{n}\bar{n}$ mode to be 
\begin{eqnarray}
\tau^{-1, Nuc}_{nn\to\bar{n}\bar{n}}\sim  \frac{\Delta E^5_{Nuc.}G^4_{nn\Phi}\epsilon^2}{m^4_{\Phi}}.
\end{eqnarray}
When two neutrons transform to two antineutrons, the $\bar{n}$'s will annihilate and emit pions. For the case of $n-\bar{n}$ oscillation, one expects 4 to 5 pions coming out. No such discusssion has been done for our case of $\Delta B=4$ change of baryon number in the literature. If we assume a similar multi-pion emission, we can take the current bound on $n-\bar{n}$ transition in nulclei as a rough guide and use  $10^{32}$ years as the lower bound on the lifetime for nuclear $\Delta B=4$ processes. We then get a limit on the parameter combination $(G^2_{nn\Phi}\epsilon/m^2_\Phi)\leq10^{-27}$ GeV$^{-2}$. Given  as noted that $G_{nn\Phi}$ is a very small number since $G_{nn\Phi} \langle \Phi\rangle$  gives rise to $n-\bar{n}$ oscillation, this gives a rather weak bound on the $\epsilon/m^2_\Phi$.


\section{10. Conclusion} In summary, we have focussed on a particular scenario for AD baryogenesis and discussed its implications for various $B$ and $L$ violating processes. 
For a given choice of the AD field mass and the associated $B/L$-violating operator, we discuss whether an observable $B$ or $L$-violation is compatible with AD baryogenesis. 
We find it interesting that a split scale scenario for AD baryogenesis is compatible with observable $n-\bar{n}$ oscillations and  the model is  embeddable into an SO(10) GUT suggested in Ref.~\cite{babumoh}. We also find a narrow range of parameters in the 100 TeV mass range fo the $\Phi, \Lambda$ and $v_\Phi$ where observability of $n-\bar{n}$ oscillation is compatible with AD baryogenesis.
We also point out that if this AD baryogenesis scenario is used for $L$-violating operators for neutrino mass, 
it either gives a neutrino mass smaller than required to fit oscillation data for atmospheric neutrinos 
or fails to satisfy the conditions for our model for  $m_\Phi < M_P$.

\section*{Acknowledgement} 
We would like to thank Bhupal Dev and Mike Wagman for discussions.
The work of R.N.M. is supported by the US National Science Foundation grant no.~PHY-1914631 and  the work of N.O. is supported by the US Department of Energy grant no.~DE-SC0012447.

\end{document}